\documentstyle[12pt]{article}

\begin{document}

\baselineskip 6mm
\renewcommand{\thefootnote}{\fnsymbol{footnote}}

\begin{titlepage}

\hfill\parbox{4cm}
{ KIAS-P99007 \\ hep-th/9901105 \\ January 1999}

\vspace{15mm}
\begin{center}
{\LARGE \bf
Non-perturbative membrane spin-orbit \\ couplings
in M/IIA theory}
\end{center}

\vspace{5mm}
\begin{center} 
Seungjoon Hyun\footnote{\tt hyun@kias.re.kr},
Youngjai Kiem\footnote{\tt ykiem@kias.re.kr}
and Hyeonjoon Shin\footnote{\tt hshin@kias.re.kr}
\\[5mm]
{\it 
School of Physics, Korea Institute for Advanced Study,
Seoul 130-012, Korea
}
\end{center}
\thispagestyle{empty}

\vfill
\begin{center}
{\bf Abstract}
\end{center}
\noindent
Membrane source-probe dynamics is investigated in the 
framework of the finite $N$-sector DLCQ $M$ theory 
compactified on a transverse two-torus for an arbitrary 
size of the longitudinal dimension.  The non-perturbative 
two fermion terms in the effective action of the matrix 
theory, the (2+1)-dimensional supersymmetric Yang-Mills 
theory, that are related to the four derivative $F^4$ terms
by the supersymmetry transformation are obtained, including 
the one-loop term and full instanton corrections.  On the 
supergravity side, we compute the classical probe action up 
to two fermion terms based on the classical supermembrane 
formulation in an arbitrary curved background geometry produced 
by source membranes satisfying the BPS condition; two fermion 
terms correspond to the spin-orbit couplings for membranes.  
We find precise agreement between two approaches when the 
background space-time is chosen to be that of the DLCQ $M$ 
theory, which is asymptotically locally Anti-de Sitter.
\vspace{2cm}
\end{titlepage}

\baselineskip 7mm
\renewcommand{\thefootnote}{\arabic{footnote}}
\setcounter{footnote}{0}

\section{Introduction}

By now considerable body of evidence toward the feasibility 
of the quantum description of $M$ theory via matrix
theory \cite{bfss,susskind} has been accumulated.   
Especially within the framework of the discrete light-cone 
quantization (DLCQ), the explicit scattering calculations
performed in matrix theory were successfully compared
to the supergravity calculations; Becker, Becker, Polchinski, 
and Tseytlin considered the scattering between two
$D$-particles ($M$-momentum) and showed that the matrix 
theory side
calculation for the effective action precisely reproduces the
eleven-dimensional supergravity side calculation up to two loops
\cite{becker}\footnote{In Ref.~\cite{add1}, an LSZ formalism
for the scattering problems in the context of the 
eleven-dimensional $M$ theory was developed.  Recently,
the general two-body scattering perturbative dynamics in
$M$ theory was systematically analyzed in Ref.~\cite{add2}
up to four fermion terms.}.  Similarly in the context of the
membrane scatterings, especially for the weak coupling
limit (the limit where the size of the longitudinal
eleventh circle is small), the agreement between the two 
approaches was obtained by many authors \cite{malda2}-\cite{tsey2}.
Recalling that the focus of the most of these analysis has been
the perturbative brane dynamics, what we attempt in 
this paper is a systematic study of the non-perturbative
brane dynamics.  

Our approach is based on two recent lines of developments.
First, it was observed in \cite{hks,hyun,hyun2} that the
appropriate space-time background geometries for the 
description of the $N$-sector DLCQ $M$ theory compactified
on a transverse $p$-torus ($p >1$) are not
asymptotically flat\footnote{For the $M$-momentum dynamics 
and the matrix
quantum mechanics, as was first formally noted 
in \cite{becker} and clarified in \cite{hks}, the background 
geometry is described by the (zero-mode part of) Aichelberg-Sexl 
type shockwave geometry \cite{hyun2}.  This geometry is 
asymptotically flat in eleven dimensions and the time coordinate
is asymptotically light-like.}.  For the membrane dynamics,
that can be most easily studied within the context
of the $M$ theory compactified on a transverse two-torus, the 
relevant background geometry is asymptotically locally
Anti-de Sitter (AdS) type \cite{hyun,hyun2}.  In this paper,
we study, in detail, the consequence to the effective action
of these non-asymptotically flat background geometries.
Second, initiated by Stern, Sethi and Paban \cite{sethi2,sethi}, 
it has been noted that the strong coupling dynamics and thus the 
effective action of supersymmetric gauge theories are 
strongly constrained by the supersymmetry.    
In the case when there are sixteen supercharges, the 
constraints are strong enough to uniquely determine
the full non-perturbative eight fermion terms of the 
effective action (up to an overall constant) 
of the (2+1)-dimensional
supersymmetric Yang-Mills (SYM) theory \cite{sethi}, which 
are related to the bosonic four derivative $F^4$ terms
by supersymmetry transformations.  The 
analysis presented in this paper gives an $M$ theory 
interpretation of the Stern-Sethi-Paban's work
in terms of the $M$ theory in non-asymptotically flat 
background geometries of Refs.~\cite{hyun,hyun2}.

According to the arguments of Seiberg and Sen \cite{seiberg},
the microscopic dynamics of DLCQ $M$ theory compactified
on a transverse two-torus is described by the 
(2+1)-dimensional
SYM theory (we loosely call it matrix theory throughout
this paper).  What we find in this paper is; the DLCQ
supergravity effective action for membrane dynamics
whose computation is based on 
Ref.~\cite{hyun,hyun2} precisely agrees with the matrix theory 
effective action calculated by the techniques based
on Ref.~\cite{sethi2,sethi} for an arbitrary value of
the coupling constant (or the size of the longitudinal
eleventh circle), consistent with Ref.~\cite{susskind}.  This 
precise agreement will be explicitly
verified for the bosonic four derivative $F^4$ terms and
the two fermion terms related to them by supersymmetry
in the effective action independently computed in both 
approaches.  It turns out that
the two fermion terms, now including full
non-perturbative corrections, can be interpreted as spin-orbit
couplings for membranes.  In the context of the matrix
quantum mechanics versus supergravity, the spin-orbit couplings
for the D-particles were successfully computed and positively 
compared in both approaches at the perturbative
level \cite{harvey}-\cite{review}.  Since we consider 
an arbitrary value of the longitudinal eleventh circle size, 
our results apply equally well to the IIA theory D2-branes, 
as well as to the eleven-dimensional $M$ membranes.  
When we take the size
of the eleventh circle to the infinity, the background
geometry of Refs.~\cite{hyun,hyun2} becomes an Anti-de Sitter
space tensored with a fixed size seven sphere, 
$AdS_4 \times S^7$.  In this case, the duality conjecture
of Maldacena \cite{malda,witten} relates the supergravity to 
the conformal phase (the infra-red limit) of the 
(2+1)-dimensional SYM theory.  Since we independently compute the 
effective action for each theory, our results constitute 
a strong supporting evidence for the Maldacena conjecture.    

The material presented in Sec.~2.1 and Sec.~2.2 is partially 
based on our results reported in Ref.~\cite{hks2}
on the membrane dynamics in the DLCQ $M$ theory.

\section{Membrane dynamics: DLCQ $M$ theory compactified on
a transverse two-torus}

This section is organized as follows.  In Sec.~2.1, 
we start from the calculation of the effective
action for the spinless probe membrane moving in the background
geometry of spinless source membranes in the DLCQ 
supergravity framework.  Since the longitudinal eleventh 
direction is also compactified as dictated by the DLCQ
prescription \cite{susskind}, our effective action includes 
all the contribution from the mirror membranes.  This
result is reshuffled by applying the Poisson resummation
formula along the eleventh direction for further
analysis.  
In Sec.~2.2, we show the correspondence
between the DLCQ supergravity and matrix theory
at the level of the bosonic effective action.
Our logic is as follows; utilizing the sixteen
supersymmetries of (2+1)-dimensional SYM theory, Stern,
Sethi and Paban \cite{sethi} determined the exact
eight fermion terms in the effective action.  We start
by recalling why their analysis works and, based on
a supersymmetry argument, we calculate
the exact four derivative bosonic effective action 
from their eight fermion terms.  Thus determined
effective action from the SYM theory is shown to be
identical to the supergravity bosonic
effective action computed in Sec.~2.1.  Based on the
same supersymmetry argument, we sketch how to recursively
determine higher fermion terms from the bosonic 
four derivative terms.  In particular, we
explicitly obtain the two fermion terms 
including the one-loop and full instanton 
corrections\footnote{In Ref.~\cite{scsqm}, we determined, 
in the framework of supersymmetric quantum mechanics, all 
fermion one-loop exact terms via supersymmetric completion
and worked out the corrected supersymmetry transformation 
due to the inclusion of the four derivative terms.}. 
In Sec.~2.3, the DLCQ
supergravity side meaning of the two-fermion terms of 
the matrix theory 
effective action is investigated.  Instead of 
considering a spinless
probe membrane, we consier the spinning probe
membrane dynamics using the curved background supermembrane 
formalism of Ref.~\cite{bst}, that was further analyzed in
Ref.~\cite{dpp}, while
for simplicity the source membranes are still kept 
spinless.  This analysis is performed on a 
general curved background geometry produced by source
membranes satisfying the BPS condition.
The leading two fermion contribution of the spin effects
to the effective action turns out to be the 
spin-orbit couplings; we explicitly determine the spin-orbit
couplings for membranes.  This two fermion effective
action obtained from the purely supergravity side analysis 
is exactly identical 
to the two fermion terms computed in matrix theory in Sec.~2.2
when we choose the harmonic function of the DLCQ supergravity
as in Sec.~2.1.  Our analysis in Sec.~2.3 points toward the 
possibility that 
the effective action computed from both approaches should agree 
for all fermion number terms. 

\subsection{Preliminary: Bosonic effective action from DLCQ 
supergravity analysis and the SYM theory basics}

Following the arguments of Seiberg and Sen to take appropriate
chains of $U$-dual transformations, the background geometry of 
the $N$-sector DLCQ $M$ theory compactified on
a transverse two-torus is given by the following 
eleven-dimensional covering space metric
\cite{hyun,hyun2}
\begin{equation}
\label{m2metric}
ds^2_{11} = h^{-2/3} (-dt^2+dx_8^2+dx_9^2)
       +h^{1/3} (dx_1^2 + \cdots + dx_7^2+dx_{11}^2) , 
\end{equation}
where the covering space eleventh coordinate $x_{11}$ parameterizes
a real line, with the periodic identification via 
\begin{equation}
 x_{11} \simeq x_{11}+ 2 \pi R .
\end{equation}
The eleventh direction thus becomes a circle with a radius $R$.
The $N$ coincident source membranes wrap the torus that extends over the
$x_8$ and $x_9$ directions.  The eleven-dimensional harmonic 
function $h$ is given by 
\begin{equation}
h= \sum_{n = -\infty}^{\infty}
 \frac{\kappa N}{(r^2+( x_{11} + 2 \pi R  n)^2 ) ^3} , 
\label{m2harmonic}
\end{equation}
where $\kappa$ is a dimensionful constant
and $r^2 = x_1^2 + \cdots + x_7^2 $ is an $SO(7)$ invariant. 
The harmonic function $h$ contains the contribution from all
mirror charges to respect the periodicity under the lattice
translation $x_{11} \rightarrow x_{11}+ 2 \pi R$.  

In the limit of the vanishingly small $R$, we can replace the summation
in Eq. (\ref{m2harmonic}) with an integration and recover the
near-horizon geometry of the $N$ $D2$-branes of the type IIA
supergravity.  At the decompactification limit of the DLCQ $M$ theory, 
that corresponds to the large $R$ limit, the eleventh direction 
becomes indistinguishable from other noncompact directions 
$( x_1 , \cdots , x_7 )$.  In particular, the summation in the 
expression for $h$ gets dominated by the $n=0$ term, which has 
the manifest $SO(8)$ symmetry; the perpendicular $SO(7)$ symmetry 
gets enhanced to the $SO(8)$ symmetry at the decompactification
limit.  In terms of an $SO(8)$ 
invariant $\tilde{r}^2 = r^2 + x_{11}^2$, the harmonic function $h$
in this limit has a simple power law dependence on $\tilde{r}$
like $\tilde{r}^{-6}$.  Since the transversal space metric
$h^{1/3}$ scales as $\tilde{r}^{-2}$, we see that the
background geometry precisely becomes $AdS_4 \times S^7$,
where the seven-sphere $S^7$ has a constant size \cite{nissan}.
This is the limit where we have the large $N$ correspondence
between the AdS supergravity and conformal field theory (CFT), 
in which the AdS supergravity and the CFT near the
infrared fixed point, i.e., the conformal phase of the
(2+1)-dimensional SYM theory, become a dual description to each 
other \cite{malda}.  

In the context of the ${\cal N} = 8$, 
(2+1)-dimensional SYM theory, the moduli 
space of the Coulomb branch is described by $N$ abelian dual magnetic 
$\phi^8$ scalars and $7N$ scalars $\phi^i$, where 
$i= (1, ... , 7)$ is the vector index of the $SO(7)$ $R$-symmetry.  
The Yang-Mills coupling constant $g_{\rm YM}^2$ has mass 
dimension one, 
and the values of the $N$ magnetic scalars should be periodically
identified with a period
proportional to $g_{\rm YM}^2$.  The moduli space is then $N$
symmetric product $S^{N} (R^7 \times S^1)$.  In our supergravity
set-up, we have $N$ identical source membranes, whose BPS solution 
space can be parametrized by $7N$ positions in the non-compact
direction $(x_1, ..., x_7)$ and $N$ positions along the
$M$ theory circle; we reserve the right to construct multi-center
solutions from Eq.~(\ref{m2harmonic}) without violating the
BPS condition.  At the origin of the SYM theory moduli space, i.e.,
in the case of the $N$ coincident source membranes as in 
Eq.~(\ref{m2harmonic}), it is known that the SYM theory flows
to an interacting $Spin(8)$ invariant theory in the infra-red
limit \cite{spin8}.  Since the $g_{\rm YM}^2$ has mass dimension
one, the infra-red limit corresponds to the strong coupling
limit.  The arguments of Seiberg and Sen \cite{seiberg}
imply $g_{\rm YM}^2
=  g_s / l_s = M_p^3 R^2$, where $g_s$, $l_s$ and $M_p$
denote the string coupling, string scale and the 
eleven-dimensional Planck mass, respectively\footnote{From
Eq.~(\ref{identi}), we find that $\phi^8 = x_{11} / l_s^2$.  
Since the period of $x_{11}$ is proportional to $R$, the 
period of $\phi^8$ should be proportional to 
$R/l_s^2 = R^2 M_p^3 = g_{\rm YM}^2$, as mentioned before.}. The
strong coupling limit in the SYM theory consequently implies 
the decompactification limit $R \rightarrow \infty$ on 
the supergravity
side.  We have already seen from Eq.~(\ref{m2metric}) that 
in the decompactification limit, the perpendicular symmetry
of the background geometry enhances from $SO(7)$ to $SO(8)$,
and the background geometry becomes $AdS_4 \times S^7$
for the $N$ coincident source membranes.  This suggests the
validity of the aforementioned duality between the infra-red, 
i.e., conformal, phase of the (2+1)-dimensional SYM and the 
$AdS_4$ supergravity.  

One of the main themes of our paper, the correspondence between
the matrix theory and the supergravity on the asymptotically
locally Anti-de Sitter background geometry is motivated by the
consideration along the above line at least for the
large $R$ limit.  Our primary interest here,
however, will be to study the case of the arbitrary values of 
$N$ and $R$ (thus $g_{\rm YM}^2$) following the DLCQ
prescription of Ref.~\cite{susskind}.  On the supergravity side,
the treatment of the finite $R$ is straightforward; we simply
have to add all contributions from the mirror membranes
to respect the lattice
translation symmetry $x_{11} \rightarrow x_{11} + 2 \pi R$,
as we did in Eq.~(\ref{m2harmonic}).
However, on the SYM theory side, we expect considerable 
instanton
corrections when $g_{\rm YM}^2$ is not very small\footnote{For
small $R$ and thus small $g_{\rm YM}^2$, we are in the regime where 
we can use the IIA supergravity analysis.  On the SYM theory side, 
perturbative calculations would be enough.  
The agreement between 
the perturbative one-loop SYM theory and the classical D2-brane 
dynamics was reported already in the 
literature \cite{malda2,tsey,lif} (modulo 
the $r$ independent
$v^4$ term which 
vanishes in the DLCQ supergravity as shown in Eq.~(\ref{appr})).}.  
The generic contributions from instantons to the effective potential
are exponential terms, while the harmonic function from the
supergravity has power law dependence.  The key observation
to solve this apparent problem is to recall that the
(2+1)-dimensional SYM theory is the dimensional reduction of the
ten-dimensional SYM theory.  We reshuffle the series summation of 
Eq.~(\ref{m2harmonic}), that is the lattice translation
along the {\em eleventh} circle, using the Poisson resummation
formula:
\begin{equation}
\sum^{\infty}_{n=-\infty} f(n) =
   \sum^{\infty}_{m=-\infty} \int^\infty_{-\infty} d \phi \;
   f( \phi ) \; e^{2 \pi i m \phi} ~.
\end{equation}
The resummation can be exactly performed to yield the following
identity.
\begin{equation}
  \sum_{n = -\infty}^{\infty}
  \frac{1}{(r^2+( x_{11} + 2 \pi R  n)^2 ) ^3} 
\label{presum}
\end{equation}
\begin{equation}
  =    \frac{1}{16 R} \ \Big[ \  \frac{3}{ r^5} + 
\sum^{\infty}_{m=1} e^{-mr/R} (\frac{m^2}{R^2} \frac{1}{ r^3}
   +   \frac{m}{R} \frac{3}{ r^4} + \frac{3}{r^5} )
     2 \cos (mx_{11}/R)~ \Big]  
     \label{resum} 
\end{equation}
\begin{equation}
  = \frac{1}{16 R} 
    \left[ \; \frac{3}{r^5} +
       \sum^{\infty}_{m=1} \left( \frac{2}{\pi} \right)^\frac{1}{2}
              \frac{m^2 m^{1/2}}{R^5} \left( 
         \frac{R}{r} \right)^{5/2}
                           K_{5/2} ( mr / R ) 
        2 \cos (m x_{11} / R ) \;
    \right] . 
\label{resum2}
\end{equation}
In going from (\ref{resum}) to (\ref{resum2}),
we use the modified Bessel function $K_\nu$ with a half-integer
$\nu$, which has the finite number of terms in an expansion
\cite{table}
\begin{equation}
\label{bessel}
K_{j+1/2} (z) = \left( \frac{\pi}{2 z} \right)^\frac{1}{2}
	e^{-z} \sum^j_{k=0} \frac{ (j+k)! }{ k! (j-k)! (2z)^k } ~.
\end{equation}
Each term of (\ref{presum}) is the harmonic function of the
eight dimensional space $(x_1, \cdots, x_7, x_{11})$
perpendicular to the source membranes.
The first term of (\ref{resum}) is in fact the harmonic
function of the seven dimensional $(x_1, \cdots , x_7)$ space.
As such, it appears in the construction of the IIA supergravity 
D2-brane solutions.  It vanishes when we act $(\partial_{11})^2$
and thus it is the contribution from the
massless modes under the Kaluza-Klein dimensional reduction 
along the $M$ theory circle.  The remaining {\em exponential} 
terms are from the massive Kaluza-Klein modes; when we act
$(\partial_{11})^2$ to the $m$-th term, we get the
eigenvalue $- m^2 / R^2$.  In Ref.~\cite{joe}, noting 
that $2 \cos x = \exp (ix) + \exp (-ix)$, these remaining terms 
were interpreted as originating from the $M$-momentum transfer 
between the source and probe D2-branes.  From the Yang-Mills
theory point of view, the exponential terms look generically
like the $m$-instanton contributions.  The $r^{-4}$ and $r^{-5}$
terms of the $m$-th term in (\ref{resum}) represent 
the perturbative
corrections in the $m$-instanton background.  A priori, these
perturbative corrections should continue to all orders
of the coupling $g_{\rm YM}^2$.  However, as we will show in 
Sec.~2.2,
the constraints from the remaining sixteen supersymmetry based 
on the argument of Ref.~\cite{sethi} cut the contribution at 
the finite order.
 
We now consider the purely bosonic dynamics of a probe membrane, 
which is taken to span the $x_8$, $x_9$ directions and 
is moving with a constant velocity 
$v^{\hat{I}}=\partial_0 x^{\hat{I}}$ 
($\hat{I}=1, \cdots, 7$) in a direction perpendicular to the probe
and $x_{11}$.  The background geometry for the finite value of $R$
has the $SO(7)$ symmetry and the velocity is an $SO(7)$
vector, consistent with the symmetry of the background geometry.  
The action for the probe
membrane is
\begin{equation}
\label{m2action}
S = T_2 \int d^3 \zeta 
\Big[ - \sqrt{-g } - \frac{1}{6}
\epsilon^{ijk} \partial_i x^{\hat{\mu}} \partial_j x^{{\hat\nu}} 
\partial_k x^{\hat{\rho}} C_{ \hat{\rho} \hat{\nu} 
\hat{\mu} } \Big] ~,
\end{equation} 
where $T_2$ is the membrane tension and 
$C_{\hat{\mu} \hat{\nu} \hat{\rho}}$ is the 
three-form gauge field of the eleven-dimensional supergravity. 
Here $i$, $j$, $k$ are the world-volume indices and the hatted indices
represent the eleven-dimensional indices. 
The metric $g_{ij}$ is the induced metric on the 
world-volume of the probe membrane given by
\begin{equation}
g_{ij} = g_{\hat{i} \hat{j}} + 
\partial_i x^{\hat{I}} \partial_j x^{\hat{J}}
 g_{\hat{I}\hat{J}}~,
\end{equation}
where the indices $\hat{I}$, $\hat{J}$ represent the directions 
perpendicular
to the probe. We choose the static gauge where
$\partial_0 x^{\hat{0}} = \partial_1 x^{\hat{1}} 
= \partial_2 x^{\hat{2}} = 1$
and other derivatives of $x^{\hat{i}}$ with respect to
$x^{j}$ are zero.
We plug the metric Eq.~(\ref{m2metric}) with the function $h$ 
of Eq.~(\ref{m2harmonic}) into the action $S$ and expand it in 
powers of the transverse velocity $v$. The action $S$ becomes
\begin{equation}
S = \int d^3x \; \left[ \; \frac{1}{2} T_2 v^2
                 - V_2 + {\cal O} ( (v^2)^3 )
               \right] ~,
\end{equation}
where $V_2$ is the effective potential given by
\begin{eqnarray}
V_2 &=& -\frac{1}{8} T_2 h(r,x_{11}) (v^2)^2 \nonumber \\
    &=& - \frac{N}{16 R M_p^{3}} (v^2)^2 \nonumber \\
    & & \times  \Bigg[ \;  \frac{3}{r^5} 
             + \sum^{\infty}_{m=1} 
               \left( \frac{2}{\pi} \right)^\frac{1}{2}
             \frac{m^2 m^{1/2}}{R^5} 
                   \left( \frac{R}{r} \right)^{5/2}
                     K_{5/2} (mr/R) 2 \cos (m x_{11}/R ) \;
        \Bigg] ~. \label{sugrapot}
\end{eqnarray}
Going to the last line, we use the fact that $T_2 \kappa = 8 M_p^{-3}$
where $M_p$ is the eleven-dimensional Planck scale \cite{joe}
and perform the Poisson resummation.  It
should be noted that the potential is valid for {\it any} value of
$R$.  If $R$ is very small (or $r \gg R$), 
 the potential is approximated by
\begin{equation}
V_2 \approx -\frac{3N}{16 R M_p^3} \frac{(v^2)^2}{r^5}
            -\frac{N}{16 R^3 M_p^3} \frac{(v^2)^2}{r^3} e^{-r/R} 
                 2 \cos (x_{11}/R)  ~.
\label{appr}
\end{equation}                                        
The first term of Eq.~(\ref{appr}) is the usual potential 
between two
D2-branes in the ten-dimensional type IIA theory \cite{lif} and
the second term is the potential due to the effect of a single
$M$-momentum transfer \cite{joe}. The approximate potential
Eq.~(\ref{appr}) shows a notable feature that there is no $r$
independent $v^4$ term that appeared in \cite{joe}.  Had we
started from an asymptotically flat background geometry,
that term will inevitably appear.  In the large
$N$ limit, it is natural to drop the term as was done in, for
example, \cite{lif}.  In the DLCQ framework, however, this term
is automatically absent \cite{tsey2,hks}.  This feature is
also present in the case of the exact 
potential, Eq.~(\ref{sugrapot}).

\subsection{Matrix theory calculation of two fermion 
terms: supersymmetric completion}

According to the prescriptions of Seiberg and Sen, the 
DLCQ $M$-theory
on a transverse two-torus is described by 
a system of D2-branes wrapped on its
$T$-dual two-torus \cite{seiberg}, which becomes
very large when the original two-torus has a vanishingly
small size. When the number of
D2-branes is $N$, the action for the system is just the
(2+1)-dimensional $U(N)$ SYM theory. The effective potential between
the source and the probe membranes is 
given by the effective potential
of the SYM theory, and we compare the supergravity bosonic
effective potential
Eq.~(\ref{sugrapot}) to the bosonic
effective potential of the SYM theory.  We
note that our supergravity side calculation is actually for the
two-body dynamics of the source and the probe.   
>From the gauge theory point of view, we do not
give the vacuum expectation values to the scalars that represent the
position of the $N$ source membranes, thereby making them localized at
one transversal space-time point, corresponding to the
origin of the SYM theory moduli space. 

Since the metric of the moduli space of the (2+1)-dimensional
SYM theory is flat, the quadratic effective action of the SYM 
theory can
be straightforwardly written as\footnote{For our later purpose,
we do not write terms in action with $\partial_1 \phi^i$
and $\partial_2 \phi^i$.  Similarly, except for the
supersymmetric partner terms of the bosonic 
quadratic terms,
we do not write fermion derivative terms.  The spinors have
{\bf 2} of $SO(2,1)$ indices and it is always implicitly assumed 
that an appropriate $2 \times 2 $ matrix is sandwiched between
two fermions.  We use {\bf 8} representation of
$Spin(7)$, but sometimes we implicitly use {\bf 8$_c$}
or {\bf 8$_s$} of $Spin(8)$.  Essentially, we are considering 
the `center of mass' dynamics of the probe membrane.  As such,
our presentation closely parallels the supersymmetric quantum
mechanics of \cite{sethi2} and we follow their notation
for the most part.}
\begin{equation}
\Gamma^{(0)} =  \int d^3 x \left( \frac{1}{2}  u^i u^i 
 + \frac{i}{2} \psi  \dot{ \psi} \right) ~,
\label{taction}
\end{equation}
where  $u^i = \dot{\phi}^i= F_{0 i}$, the $i$-th component of the
electric field. The scalars $\phi^i$ ($i=1, \cdots , 7$) are the 
seven scalars of the vector multiplet (thereby having the $SO(7)$ 
symmetry).  Assigning an ordering $O(\partial_{\mu}) = 1$ and
$O(\psi) = 1/2$, we note that 
$O(\Gamma^{(0)}) = 2$.  The action (\ref{taction}) is invariant 
under the tree-level supersymmetry transformation:
\begin{equation}
 \delta \phi^i =  -i \epsilon \gamma^i \psi ~ ,
\label{tsusy}
\end{equation}
\[ \delta \psi  = \epsilon \gamma^i  \frac{d}{dt} 
 \phi^{i}  =   u^i \epsilon \gamma^i  ~, \]
where we assign $O(\epsilon)= -1/2$. 
The general structure of the effective
action $\Gamma^{(1)}$, which is of the 
order $O(\Gamma^{(1)})= 4$, looks schematically
like \cite{harvey,morales}
\begin{equation}
 \Gamma^{(1)} =  \int d^3 x \left( f^{(0)} u^4 
+ f^{(2)} u^3 \left[ \psi^2 \right] 
+ f^{(4)} u^2 \left[ \psi^4 \right]
+ f^{(6)} u \left[ \psi^6 \right]
+ f^{(8)}  \left[ \psi^8 \right] \right) , 
\label{scheme}
\end{equation}
where $\left[ \psi^p \right] $ denotes a generic $p$ fermion
structure, and $f^{(p)}$ represents the bosonic 
coefficient function
of the corresponding $p$ fermion structure.  Upon adding
$\Gamma^{(1)}$ to the quadratic effective action
$\Gamma^{(0)}$, the supersymmetry transformation law
in Eq.~(\ref{tsusy}) should be modified; we thus
write the $\Gamma^{(1)}$-corrected supersymmetry 
transformation as
\begin{equation}
 \delta \phi^i =  -i \epsilon \gamma^i \psi  + \epsilon N^i \psi ,
\label{susy}
\end{equation}
\[ \delta \psi  = u^i \epsilon \gamma^i  +  M \epsilon ~. \]
We note that $O(N)=2$ and $O(M)=3$ and, therefore, we
can schematically write 
\begin{equation}
N^i = u^2 N^{i(0)} + u N^{i(2)} \left[ \psi^2 \right]
   +  N^{i(4)} \left[ \psi^4 \right]
\label{Nexp}
\end{equation}
and
\begin{equation}
M = u^3 M^{(0)} + u^2 M^{(2)}  \left[ \psi^2 \right]
+  u M^{(4)}  \left[ \psi^4 \right]
 +  M^{(6)}  \left[ \psi^6 \right] ~.
\label{Mexp}
\end{equation}
The effective action $\Gamma^{(0)} + \Gamma^{(1)}$
should be invariant under the supersymmetry 
transformation Eq.~(\ref{susy}) order by order.
At the lowest order, $O = 2$, $\Gamma^{(0)}$ itself
is invariant under Eq.~(\ref{tsusy}).  At the next
order, $O= 4$, we have two contributions; one
from the supersymmetry transformation of $\Gamma^{(0)}$
due to the corrections $N^i$ and $M$, and another
from the variation of $\Gamma^{(1)}$ under 
Eq.~(\ref{tsusy}).  
The variation $\delta ( \Gamma^{(0)} + \Gamma^{(1)} )$ contains
one, three, five, seven and nine $\psi$ terms, and they have
to separately vanish for the invariance of the effective
action under supersymmetry transformations.
Specifically, we have: 
\begin{equation}
\delta_B ( f^{(0)} u^4 ) 
+ \delta_F (  f^{(2)} u^3 \left[ \psi^2 \right] )
+ u^i u^2 \epsilon \dot{N}^{i(0)} \psi
+ \frac{i}{2} u^3 \psi  \dot{M}^{(0)} \epsilon
=0 ~,
\label{onef}
\end{equation}
\begin{equation}
\delta_B (  f^{(2)} u^3 \left[ \psi^2 \right]   ) 
+ \delta_F (  f^{(4)} u^2 \left[ \psi^4 \right]  )  
+ u^i u \epsilon \dot{N}^{i(2)} \left[ \psi^2 \right] \psi
+ \frac{i}{2} u^2 \psi \dot{M}^{(2)} 
 \left[ \psi^2 \right] \epsilon
=0 ~,
\label{threef}
\end{equation}
\begin{equation}
\delta_B (  f^{(4)} u^2 \left[ \psi^4 \right]  ) 
+ \delta_F ( f^{(6)} u \left[ \psi^6 \right] )  
+ u^i  \epsilon \dot{N}^{i(4)} \left[ \psi^4 \right] \psi
+ \frac{i}{2} u \psi \dot{M}^{(4)} 
 \left[ \psi^4 \right] \epsilon
=0 ~,
\label{fivef}
\end{equation}
\begin{equation}
\delta_B ( f^{(6)} u \left[ \psi^6 \right] ) 
+ \delta_F (  f^{(8)}  \left[ \psi^8 \right]  )  
+ \frac{i}{2}  \psi \dot{M}^{(6)} 
 \left[ \psi^6 \right] \epsilon
=0 ~,
\label{sevenf}
\end{equation}
\begin{equation}
\delta_B (   f^{(8)}  \left[ \psi^8 \right]    )  
=0 ~, 
\label{ninef}
\end{equation}
where $\delta_B$ and $\delta_F$ represent the supersymmetric 
variation of the bosonic fields and the fermionic
fields, respectively.
The key insight of Paban, Sethi and Stern \cite{sethi} is that 
the eight fermion terms can be exactly computed via 
Eq.~(\ref{ninef}) up to an overall constant {\em without} 
the knowledge of other fermion number terms, $N^i$ and $M$;
$[ \psi^8 ]$ consists of terms with zero, two and four scalar
structure.  The scalar number $s$ ($s=0,2,4$) represents
the number of scalar fields contracted to the fermion
structure.  Generically, the effective action $\Gamma^{(1)}$ 
of the SYM theory is
the summation of a perturbative term and $m$-instanton terms.
Hereafter, for the notational convenience, we call the perturbative
term $0$-instanton sector. 
The function $f^{(p)}$
consists of the instanton summation and we represent it as $f^{(p)} =
\sum_{m=-\infty}^{\infty} f_m^{(p)} $ where $m$ is the instanton
number.  
>From \cite{sethi}, $f^{(8)}_{4,m}$ is given by
\begin{equation}
f^{(8)}_{4,m} =  m^6 |m|^{1/2} \frac{1}{g_{\rm YM}^{28}} 
	  \left( \frac{g_{\rm YM}^2}{\phi} \right)^{13/2}
	  K_{13/2} (|m| \phi / g_{\rm YM}^2 ) 
          e^{i m \phi^8 /g_{\rm YM}^2 } ~,
\label{8f}
\end{equation}
up to an overall multiplicative constant, where the extra scalar 
$\phi^8$ is the dual
magnetic scalar.  Here, $f^{(8)}_{4, m}$ is 
the coefficient function of the four scalar
structure term among $\left[ \psi^8 \right]$ in 
the $m$-instanton sector
\begin{equation}
 \sum_{m= -\infty}^{\infty}
 f^{(8)}_{4,m} (\phi^i \phi^j \phi^k \phi^l T^{ijkl}) ,
\end{equation}
where $T^{ijkl}$ is the eight fermion structure.
The function $f^{(p)}_{q,m}$ denotes the bosonic coefficient
function of the $q$ scalar structure term of the $m$-instanton sector
in $p$ fermion terms. We remark that Eq.~(\ref{8f}) gives 
the perturbative term, when we set $m=0$, proportional to 
$\phi^{-13}$.

A crucial observation first made in Ref.~\cite{hks2} is that
the bosonic zero fermion term $f^{(0)}_m$ can also
be determined {\em without} the knowledge of 
$N^i$ and $M$, once the $f^{(8)}_{4,m}$ terms are determined.
To explicitly see this, we pick out the maximum scalar
structure term from each $2p$-fermion term
of the effective action (\ref{scheme}) ($p=0,1,2,3$):
\begin{equation}
f^{(0)}_m (u^2)^2 \ , \
f^{(2)}_{1,m} \phi_i u_j u^2 \left[ \psi^2 \right]_{ij} \ , \
f^{(4)}_{2,m} \phi_i \phi_j u_k u_l 
\left[ \psi^4 \right]_{ijkl} \ , \
\label{tt1}
\end{equation}
\[ f^{(6)}_{3,m} \phi_i \phi_j \phi_k u_l 
 \left[ \psi^6 \right]_{ijkl} \ , \
f^{(8)}_{4,m} \phi_i \phi_j \phi_k \phi_l 
\left[ \psi^8 \right]_{ijkl} . \]
Here, the functions $f^{(2p)}_{p,m}$ depend only
on an $SO(7)$ invariant $\phi^2 = \phi_i \phi_i$. 
The supersymmetric variation of the fermion fields of 
(\ref{tt1}) will contribute to one, three, five and seven
fermion terms shown in Eqs.~(\ref{onef})-(\ref{sevenf}):
\begin{equation}
f^{(2)}_{1,m} \phi_i u_j u_k u^2 
   \left[ \psi \gamma^k \epsilon \right]_{ij} \ , 
\label{tt6}
\end{equation}
\begin{equation}
 f^{(4)}_{2,m} \phi_i \phi_j u_k u_l u_m
   \left[ \psi^3 \gamma^m \epsilon \right]_{ijkl}  \ , 
\label{tt7}
\end{equation}
\begin{equation}
 f^{(6)}_{3,m} \phi_i \phi_j \phi_k u_l u_m
   \left[ \psi^5 \gamma^m \epsilon \right]_{ijkl}  \ , 
\label{tt8}
\end{equation}
\begin{equation}
 f^{(8)}_{4,m} \phi_i \phi_j \phi_k \phi_l u_m
   \left[ \psi^7 \gamma^m \epsilon \right]_{ijkl} \ , 
\label{tt9}
\end{equation}
respectively. The supersymmetric variation of the
bosonic coefficient functions of (\ref{tt1}) gives the 
following contributions to one, three, five and seven 
fermion terms shown in Eqs.~(\ref{onef})-(\ref{sevenf}):
\begin{equation}
\left( \frac{d}{\phi d \phi} f^{(0)}_m \right) 
   \phi_i ( u^2 )^2 
 (\epsilon \gamma^i \psi ) \ ,
\label{tt2}
\end{equation}
\begin{equation}
\left( \frac{d}{\phi d \phi} f^{(2)}_{1,m} \right)
\phi_i \phi_j u_k u^2 
 (\epsilon \gamma^i \psi ) \left[ \psi^2 \right]_{jk} ~,
\label{tt3}
\end{equation}
\begin{equation}
\left( \frac{d}{\phi d \phi} f^{(4)}_{2,m} \right)
   \phi_i \phi_j \phi_k u_l u_m
     (\epsilon \gamma^i \psi ) \left[ \psi^4 \right]_{jklm}~,
\label{tt4}
\end{equation}
\begin{equation}
\left( \frac{d}{\phi d \phi} f^{(6)}_{3,m} \right)
 \phi_i \phi_j \phi_k \phi_l u_m
  (\epsilon \gamma^i \psi ) \left[ \psi^6 \right]_{jklm} \ . 
\label{tt5}
\end{equation}
The supersymmetric variation of the bosonic fields
$\phi_i$'s appearing in Eq.~(\ref{tt1}) will reduce
the scalar number, and these terms are not
shown in Eqs.~(\ref{tt3})-(\ref{tt5}) since they 
are no longer maximum scalar structure terms.  
At each $p$-fermion term, there are contributions
from the maximum scalar number terms of $N^{i (p-1)}$ and $M^{(p-1)}$. 
However due to the time derivative 
\[ \frac{d}{dt} = u^i \phi^i \frac{d}{\phi d \phi} , \]
these contributions always include $u^i \phi^i$ factor.
In contrast, in (\ref{tt2}) the contribution is $\phi^i
(u^2)^2$.  In (\ref{tt3})-(\ref{tt5}), recalling that
the $2p$-fermion structure is in general a $p$-copy product of
$\psi \gamma^{i_1 j_1} \psi \cdots
 \psi \gamma^{i_p j_p} \psi$, $\phi^n$ and $u^p$ appearing 
there are always anti-symmetrized.  Therefore, the
contributions (A) from $N^{i (p-1)}$ and $M^{(p-1)}$ 
do not mix with the contributions (B) 
from (\ref{tt2})-(\ref{tt5}).
Two linearly-independent contributions (A) and (B)
should separately cancel the 
contributions from (\ref{tt6})-(\ref{tt9}) 
in Eqs.~(\ref{onef})-(\ref{sevenf}).
Thus, the function $f^{(0)}_m$ is related to $f^{(8)}_{4, m}$ by
\begin{equation}
C_p \left( \frac{d}{\phi d \phi} \right) f^{(2p)}_{p,m} = 
 f^{(2p+2)}_{p+1,m} ~~ \rightarrow ~~
 k \left( \frac{d}{\phi d \phi} \right)^4 f^{(0)}_m =  
   f^{(8)}_{4,m} \  , 
\label{relat}
\end{equation}
where $C_p$ are the numbers determined by working out the spinor 
algebra and $k= C_0 C_1 C_2 C_3 $.
Noting \cite{table} \begin{equation} \left( \frac{d}{z dz} \right)^a
(z^{-\nu} K_\nu (z)) = (-1)^a z^{-\nu -a} K_{\nu+a} (z) ~,
\end{equation}
we conclude 
\begin{equation}
\label{nonpert}
f^{(0)}_m = C  m^2 |m|^{1/2} \frac{1}{g_{\rm YM}^{12}} 
	  \left( \frac{g_{\rm YM}^2}{\phi} \right)^{5/2}
	  K_{5/2} (|m| \phi / g_{\rm YM}^2) 
          e^{i m \phi^8 /g_{\rm YM}^2 }
\label{f0}
\end{equation}
from Eq.~(\ref{relat}), where $C$ is an overall constant.  
When integrating Eq.~(\ref{relat}), there are 
in general four constants
of integration.  All these contributions, however, do not
contain exponential functions and, thus, comparing to
the well-behaved perturbative results for the weak coupling
limit calculations \cite{joe}, they are all set to zero.
The constant $C$ can not be determined by the argument so far, but 
the one-instanton calculation of Ref.~\cite{joe} determines it 
to be $C=
N (2/\pi)^{1/2} g_{\rm YM}^2/16$.  Thus, the bosonic  
effective action $\Gamma^{(1)}_B$ from the SYM theory is
\begin{equation}
\Gamma^{(1)}_B =   \int d^3x \sum_{m= -\infty}^{\infty}
 f^{(0)}_m (u^2)^2  \  , 
\label{symeff}
\end{equation}
including the full non-perturbative instanton corrections.  Since
$\Gamma^{(1)}_B = - \int d^3 x V_B$, the effective potential
$V_B$ is
\begin{eqnarray}
V_B &=& - \frac{N}{16} (u^2)^2  \nonumber \\
        & \times & \left[ \; \frac{3}{\phi^5} +
               \sum^{\infty}_{m=1} 
                     \left( \frac{2}{\pi} \right)^\frac{1}{2}
                     \frac{m^2 m^{1/2}}{g_{\rm YM}^{10} }
             \left( \frac{g_{\rm YM}^2}{\phi}
                     \right)^{5/2} 
                     K_{5/2} ( m \phi / g_{\rm YM}^2 ) 
    2 \cos (m \phi^8 / g_{\rm YM}^2 ) \;
        \right] . \nonumber \\
  &  &\label{sympot}
\end{eqnarray}
We note that Eq.~(\ref{sympot}) is exactly identical to
Eq.~(\ref{sugrapot}) if we identify 
\begin{equation}
 \phi_i = x_i / l_s^2 ~,~ \phi^8 =
x_{11} / l_s^2  ~,~ u = v / l_s^2 ,
\label{identi}
\end{equation}
and use $g_{\rm YM}^2 = g_s / l_s$.
The string coupling constant $g_s$ and the string length scale $l_s$
are related to the $M$ theory quantities by $g_s = (RM_p )^{3/2}$ and
$l_s = ( RM_p^3 )^{-1/2}$.

We now turn to the case of two fermion terms in $\Gamma^{(1)}$, which
is usually interpreted as the spin-orbit interaction.  Generally,
we can write it down as
\begin{equation}
\Gamma^{(1)}_{\rm spin-orbit} =  \int d^3 x \sum^{\infty}_{m=-\infty}
                     f^{(2)}_{1,m} u^2 u^i \phi^j 
    (\psi \gamma^{ij} \psi)~.
\label{symso}
\end{equation}
>From Eq.~(\ref{onef}), relating different scalar coefficient
functions $f^{(p)}$ and recalling our previous remarks in this
section, the function $f^{(2)}_{1,m}$ can be easily determined
from the given purely bosonic 
coefficient function $f^{(0)}_m$ in Eq.~(\ref{f0}).
Working out the simple spinor algebra in 
Eq.~(\ref{tt6}), $f^{(2)}_{1,m}$ is related to $f^{(0)}_m$ by
\begin{equation}
f^{(2)}_{1,m} =  \frac{i}{2} \left( \frac{ d }{\phi d \phi} \right)
 f^{(0)}_m  ~,
\end{equation}
or in other words
\begin{equation}
\phi_j f^{(2)}_{1,m} =  \frac{i}{2} \partial_j  f^{(0)}_m . 
\end{equation}
Written explicitly, 
the effective potential $V_{\rm spin-orbit}$ from the
two fermion terms that 
satisfies $\Gamma_{\rm spin-orbit}^{(1)} = - \int d^3 x 
V^{(1)}_{\rm spin-orbit}$ is thus
\begin{eqnarray}
V_{\rm spin-orbit} 
   &=& i \frac{N}{32} u^2 u^i \phi^j (\psi \gamma^{ij} \psi) 
                                        \nonumber \\ 
   & \times & \left[ \; \frac{15}{\phi^7} +
        \sum^{\infty}_{m=1} 
             \left( \frac{2}{\pi} \right)^{1/2}
             \frac{ m^{7/2}}{g_{\rm YM}^{14} } 
             \left( \frac{g_{\rm YM}^2}{\phi} \right)^{7/2} 
              K_{7/2} ( m \phi / g_{\rm YM}^2 ) 
             2 \cos (m \phi^8 / g_{\rm YM}^2 ) 
              \; \right]~ . \nonumber \\
   & &
\label{sopot}
\end{eqnarray}
If we rewrite the $V_{\rm spin-orbit}$ in terms of the $M$ theory
quantities, it becomes
\begin{eqnarray}
V_{\rm spin-orbit} 
   &=& i \frac{N}{32(RM^3_p)^3} v^2 v^i x^j (\psi \gamma^{ij} \psi) 
                                       \nonumber \\ 
   & & \times  \left[ \; \frac{15}{r^7} +
        \sum^{\infty}_{m=1} 
             \left( \frac{2}{\pi} \right)^{1/2}
             \frac{ m^{7/2}}{R^7} 
             \left( \frac{R}{r} \right)^{7/2} 
              K_{7/2} ( m r/R) 
             2 \cos (m x_{11} / R ) 
              \; \right]         \nonumber \\
   &= & i \frac{3N}{R^2 M^9_p} \sum^{\infty}_{n=-\infty}
         \frac{v^2 v^i x^j (\psi \gamma^{ij} \psi)}{ 
                  \left[ r^2 + (x_{11} + 2 \pi R n )^2 \right]^4 }~,
\label{sopot2}
\end{eqnarray}
where we Poisson-resummed back the expression going from the
first line to the second line.

\subsection{Membrane spin-orbit coupling from supergravity: matrix 
theory-supergravity correspondence for two fermion terms}

We now calculate membrane spin-orbit 
couplings from the classical supergravity side.  For this
purpose, we consider the dynamics of a spinning probe
membrane moving in the background geometry produced 
by spinless source membranes.  The BPS background fields
produced by the source membranes are known to be
determined by a harmonic function in IIA supergravity or
in eleven-dimensional supergravity.  A notable technical
feature of our calculation is that we perform the calculation
for an arbitrary choice of the harmonic function in the
eleven-dimensional supergravity.  Thus, by linearly superposing 
all mirror brane contributions,
which results from the compactification of the $M$ theory
circle, our results are applicable to both type IIA D-membranes
and M-membranes.  This will be useful for the comparison
to the matrix theory side calculations in Sec.~2.2, since 
we included full non-perturbative 
instanton corrections when computing the matrix theory side
results.  By appropriately choosing the constant of
motion for the harmonic function, the spin-orbit couplings
for the asymptotically flat and $SO(1,2) \times SO(8)$ invariant 
background 
geometry can be immediately written down from our analysis.  
For the precise agreement with the matrix theory side calculations,
we need, however, a non-asymptotically flat background
geometry that is asymptotically locally $AdS_4 \times S^7$.
  
In the superspace formalism with superspace coordinates 
$Z^M (\zeta ) = (X^{\mu} (\zeta) , \theta^{\alpha} (\zeta) )$
as functions of the world-volume coordinates $\zeta^i$, 
the probe dynamics of the supermembranes in the 
eleven-dimensional supergravity is 
described by the following action \cite{bst}\footnote{Our 
conventions
for indices are as follows.  We use $(\mu \nu \rho \cdots )$
for bosonic curved space indices and $(\alpha \beta \gamma
\cdots )$ for fermionic curved space indices.  
We write these two indices collectively as $(MNP \cdots )$ .
Among the bosonic indices, the directions tangential
to membranes will be denoted as $(\hat{i} \hat{j} \hat{k} 
\cdots )$, and the directions perpendicular to the 
branes, $(\hat{I} \hat{J} \hat{K} \cdots )$.  
Turning to the tangent space, we use $(rst \cdots)$ for 
bosonic tangent space indices
and $(abc \cdots)$ for fermionic tangent space indices.  
Collectively these two indices will be written as
$(ABC \cdots)$.  Among the tangent space bosonic indices,
$(\tilde{i} \tilde{j} \tilde{k} \cdots )$ represent the
directions tangential to membranes, and $(\tilde{I} \tilde{J} 
\tilde{K} \cdots )$, the perpendicular directions.  
The bosonic world volume indices will be denoted as
$(ijk \cdots)$.  Our signature choice for the metric
throughout this paper is $(-++ \cdots)$, and the totally
anti-symmetric three-form tensor satisfies 
$\epsilon_{012} = - \epsilon^{012} = +1$.}
\begin{equation}
S [ Z(\zeta ) ] = T_2 \int d^3 \zeta 
\Big[ - \sqrt{-g (Z(\zeta ) )} - \frac{1}{6}
\epsilon^{ijk} \Pi_i^A \Pi_j^B \Pi_k^C B_{CBA}
  ( Z(\zeta ) ) \Big] ,
\label{superm}
\end{equation}
where the pull-back $\Pi_i^A$ of the supervielbein
$E_{M}^{A}$ to the membrane world-volume satisfies 
$\Pi_i^A = \partial Z^M / \partial \zeta^i E_M^A$, and 
$B_{MNP}$ represents the anti-symmetric tensor
gauge superfield.  The induced metric $g_{ij}$ on the
world-volume satisfies $g_{ij} = \Pi_i^r \Pi_j^s 
\eta_{rs}$, where $\eta_{rs}$ is the Lorentz invariant
constant metric.  For a given background geometry,
we have to expand the action Eq.~(\ref{superm}) to
the quadratic terms in the Majorana spinor variable $\theta$,
which represents the probe spin.
>From Ref.~\cite{dpp}, we have the following explicit covariant 
expressions for the superfields in terms of the component 
fields up to the quadratic terms in the fermionic $\theta$:
\begin{equation}
 \Pi_i^r = \partial_i X^{\mu} (e_{\mu}^r 
-\frac{1}{4} \bar{\theta} \Gamma^{rst} \theta
\hat{\omega}_{\mu st } + \bar{\theta} \Gamma^r 
T_{\mu}^{\nu \rho \sigma \lambda} \theta 
\hat{F}_{\nu \rho \sigma \lambda} )
+ \bar{\theta} \Gamma^r \partial_i \theta + \cdots ~,
\label{a1}
\end{equation}
\begin{equation}
- \frac{1}{6} \epsilon^{ijk} \Pi_i^A \Pi_j^B \Pi_k^C B_{CBA}
= \frac{1}{6} dX^{\mu \nu \rho} 
\big[ C_{\mu \nu \rho} + \frac{3}{4}
 \bar{\theta} \Gamma_{rs} \Gamma_{\mu \nu} \theta
\hat{\omega}_{\rho}^{rs} - 3  \bar{\theta} \Gamma_{\mu \nu} 
T_{\rho}^{\sigma \lambda \kappa \tau} \theta 
\hat{F}_{\sigma \lambda \kappa \tau} \big]
\label{a2}
\end{equation}
\[ - \frac{1}{2} \epsilon^{ijk} \bar{\theta} \Gamma_{\mu \nu}
\partial_k \theta \partial_i X^{\mu} \partial_j X^{\nu}
 + \cdots . \]
The Dirac conjugate is defined as
$\bar{\theta} = i \theta^T \Gamma^{\tilde{0}}$.
Since we are considering spinless
background geometries, the background gravitino
field is set to zero.  The spin connection and the 
four-form gauge field strength for the background
geometry are denoted as $\hat{\omega}_{\mu st }$ and
\[ \hat{F}_{\mu \nu \rho \sigma}
 = 4 \partial_{ [ \mu} C_{\nu \rho \sigma ] }  , \] 
respectively, where the bracket implies the
antisymmetrization normalized to unity.
The $k$ eleven-dimensional gamma matrix products
$\Gamma^{r_1 \cdots r_k}$ are totally antisymmetrized
(normalized to unity) with respect to all 
indices.  The symbols 
$T_{\mu}^{\nu \rho \sigma \lambda}$ and $dX^{\mu \nu \rho}$
are defined as 
\[ T_{\mu}^{\nu \rho \sigma \lambda}  = 
 \frac{1}{288} ( \Gamma_{\mu}^{\nu \rho \sigma \lambda} 
- 8 \delta_{\mu}^{[ \nu}
\Gamma^{\rho \sigma \lambda ]} )    , \]
\[  dX^{\mu \nu \rho} = \epsilon^{ijk}
\partial_i X^{\mu} \partial_j X^{\nu}
\partial_k X^{\rho}  . \] 

A spinless BPS background geometry produced by
source membranes has the following metric and the gauge 
field
\begin{equation}
 ds^2 = h^{-2/3} \eta_{\hat{i} \hat{j} }
dx^{\hat{i}} dx^{\hat{j}} + h^{1/3}
\delta_{\hat{I} \hat{J} } dx^{\hat{I}} dx^{\hat{J}} ,
\label{met}
\end{equation}
\begin{equation}
C_{\hat{i} \hat{j} \hat{k} } = - \frac{1}{h}
 \epsilon_{\hat{i} \hat{j} \hat{k} } ,
\label{gauge}
\end{equation}
where $h$ is a harmonic function defined on the
transversal space to the source membranes.  
As far as the BPS condition is not violated, we
can linearly-superpose the harmonic function from
each source membrane.  The metric (\ref{met})
determines the non-vanishing vielbeins and spin
connections as 
\begin{equation}
 e_{\hat{i}}^{\tilde{j}} =
   h^{-1/3} \delta_{\hat{i}}^{\tilde{j}} 
\ \ , \ \ 
 e_{\hat{I}}^{\tilde{J}} =
   h^{1/6} \delta_{\hat{I}}^{\tilde{I}} 
\label{vielbein}
\end{equation}
\begin{equation}
 \hat{\omega}_{\hat{i}~\tilde{I}}^{~\tilde{i}} =
  \frac{1}{3} h^{-2/3} \partial_{\hat{I}} h  ~, \
 \hat{\omega}_{\hat{I}}^{~\tilde{I} \tilde{J}} =
 - \frac{1}{6} h^{-2/3} \partial_{\hat{J}} h ~,
\label{spinc}
\end{equation}
\[  \hat{\omega}_{\hat{i} \tilde{I}}^{~~\tilde{i}} = 
 - \frac{1}{3} h^{-2/3} \partial_{\hat{I}} h ~, \
  \hat{\omega}_{\hat{I}}^{~\tilde{J} \tilde{I}} = 
  \frac{1}{6} h^{-2/3} \partial_{\hat{J}} h  ~. \]
We note that the repeated indices in Eq.~(\ref{spinc})
are not summed.  For the description of the probe
membrane, we use the static gauge where we set 
$\partial_i X^{\hat{j}} = \delta_i^{\hat{j}}$.  
Due to the existence of the $\kappa$-symmetry for the
membrane action (\ref{superm}), the fermions $\theta$
are constrained to satisfy the $\kappa$-symmetry
gauge fixing condition 
\begin{equation}
 (1 - \tilde{\Gamma} ) \theta = 0 \ ,
\label{kappa}
\end{equation}
where $\tilde{\Gamma} = \Gamma^{\tilde{0} \tilde{1} \tilde{2}}$.
Paying attention to the center of mass motion of membranes,
we set 
\begin{equation}
 \partial_0 X^{\hat{I} } = v^{\hat{I}}
 \ , \ \partial_1 X^{\hat{I}} = 0 \ , \
  \partial_2 X^{\hat{I}} = 0 .
\label{moving}
\end{equation}
The static limit is when $v^{\hat{I}} = 0$.
By plugging Eqs.~(\ref{vielbein})-(\ref{moving})
into the first term of Eq.~(\ref{superm}), via
Eq.~(\ref{a1}), we obtain
\begin{equation}
 \int d^3 \zeta - \sqrt{-g (Z(\zeta ) )} =
\int - h^{-1} \sqrt{ 1 - hv^2 }
\Big[ 1 + \frac{1}{2} h^{2/3} \big[ 2 h^{-1/3} 
\bar{\theta} \Gamma^{\tilde{1}}
 \partial_1 \theta +  2 h^{-1/3} \bar{\theta} \Gamma^{\tilde{2}}
 \partial_2 \theta 
\label{temp1}
\end{equation}
\[  - \frac{1}{1-hv^2}
( - h^{-4/3} v^{\hat{I}} \partial_{\hat{J}} h
\bar{\theta} \Gamma^{\tilde{0}~\tilde{J}}_{~\tilde{I}}
\theta -2 h^{-1/3} \bar{\theta} \Gamma^{\tilde{0}}
 \partial_0 \theta +2 h^{1/6} v^{\hat{I}} \bar{\theta} 
\Gamma_{\tilde{I}}
 \partial_0 \theta ) 
   \big] + \cdots \Big] d^3 \zeta \]
up to two fermion terms.  Here $v^2$ denotes
$v^2 \equiv \delta_{\hat{I} \hat{J}} v^{\hat{I}} v^{\hat{J}} $.  
Likewise, the second term of Eq.~(\ref{superm}), via
Eq.~(\ref{a2}), is computed to be
\begin{equation}
\int d^3 \zeta - \frac{1}{6} \epsilon^{ijk} 
\Pi_i^A \Pi_j^B \Pi_k^C B_{CBA}
= \int \Big[ h^{-1}  - h^{-2/3} \bar{\theta}
(\Gamma^{\tilde{0}} \partial_0  + \Gamma^{\tilde{1}} \partial_1 
+ \Gamma^{\tilde{2}} \partial_2 ) \theta
\label{temp2}
\end{equation}
\[  + \frac{1}{2} h^{-5/3} v^{\hat{I}} \partial_{\hat{J}} h
\bar{\theta} \Gamma^{0~\tilde{J}}_{~\tilde{I}} \theta
+ h^{-1/6} v^{\hat{I}} \bar{\theta} (
\Gamma_{\tilde{I} \tilde{1}} \partial_2 - 
\Gamma_{\tilde{I} \tilde{2}} \partial_1) \theta + 
  \cdots \Big] d^3 \zeta \]
up to two fermion terms.  In deriving Eqs.~(\ref{temp1})
and (\ref{temp2}), we use the Majorana properties
for the spinor $\theta$ such as $\bar{\theta} 
\Gamma^{\tilde{r}_1 \cdots \tilde{r}_k} \theta = 0$
for $k= 1,2,5,6,9, 10$ and the $\kappa$-projection
condition (\ref{kappa}).  

For the slow speed expansion, we introduce an ordering
where $O(v) = 1$, $O(\partial_i ) = 1$ and 
$O(\bar{\theta} \theta)= 1$.
Adding Eqs.~(\ref{temp1}) and (\ref{temp2}) and retaining
terms of order up to four, we obtain
\begin{equation}
S = T_2 \int d^3 \zeta ( {\cal L}^{(2)} + {\cal L}^{(3)}
 + {\cal L}^{(4)} \cdots )
\label{sugrae}
\end{equation}
where
\begin{equation}
{\cal L}^{(2)} = \frac{1}{2} v^2 - 2 h^{-2/3}
\bar{\theta} ( \Gamma^{\tilde{0}} \partial_0 
+   \Gamma^{\tilde{1}} \partial_1 
+  \Gamma^{\tilde{2}} \partial_2 ) \theta ,
\label{gamma2}
\end{equation}
\begin{equation}
{\cal L}^{(3)} = h^{-1/6} v^{\hat{I}} \bar{\theta}
( \Gamma^{\tilde{I}} \partial_0
 + \Gamma_{\tilde{I} \tilde{1}} \partial_2
 - \Gamma_{\tilde{I} \tilde{2}} \partial_1 ) \theta  ,
\label{gamma3}
\end{equation}
\begin{equation}
{\cal L}^{(4)} = \frac{1}{8} h (v^2 )^2 
- \frac{1}{4} h^{-2/3}  v^2 v^{\hat{I}}
\partial_{\hat{J}} h \bar{\theta} 
\Gamma^{\tilde{0}~\tilde{J}}_{~\tilde{I}} \theta
+ \frac{1}{2} h^{1/3} v^2 \bar{\theta}
( \Gamma^{\tilde{1}} \partial_1 
+  \Gamma^{\tilde{2}} \partial_2  
- \Gamma^{\tilde{0}} \partial_0 ) \theta   .
\label{gamma4}
\end{equation}
Upon deleting all the two fermion terms, 
we recover the bosonic effective action
of Sec.~2.1. We note that the static potential vanishes 
up to two fermion terms consistent with the analysis of
\cite{vijay}, and the fermion terms other than the 
spin-orbit coupling term of ${\cal L}^{(4)}$ all contain
spinor field derivatives.  Since $\theta$ is a Majorana spinor
satisfying $f^2 \bar{\theta} \Gamma^{\tilde{i}} \partial_i
\theta = (f \bar{\theta}) \Gamma^{\tilde{i}}
\partial_i ( f \theta ) $
for an arbitrary scalar function $f$,
the  transformation of the spinor $\theta$ into $\psi$ via
\begin{equation}
\psi = 2h^{-1/3} \theta
\label{ftrans}
\end{equation}
brings the quadratic terms ${\cal L}^{(2)}$ to the
quadratic action (\ref{taction})
of the (2+1)-dimensional SYM theory 
with the standard normalization, recalling 
$(\Gamma^{\tilde{0}})^2=-1$.
To compare the action Eq.~(\ref{sugrae})
to the one derived in Sec.~2.2, we decompose
$SO(1,10)$ spinor $\psi$ into $SO(1,2) \times Spin(7)$
(or $SO(1,2) \times Spin(8)$ in the decompactifcation
limit $R = \infty$)
by assigning it an $SO(1,2)$ index $\alpha$ and $Spin(7)$
index (or $Spin(8)$ index in the decompactification limit)
$a$, $\psi_{\alpha a}$.  Furthermore, as a simple
background choice as before, we suppose $v^{\hat{I}}$ and
$\psi_{a \alpha}$ are a constant number and a constant
spinor, respectively.  Then, the fermion derivative
terms drop out and we finally obtain
\begin{equation}
S = T_2 \int d^3 \zeta ( \frac{1}{2} v^2  +
\frac{1}{8} h (v^2)^2 + \frac{i}{16} v^2 v^{\hat{I}}
 \partial_{\hat{J}} h \psi^{\alpha}_{a} 
   (\gamma^{\tilde{I} \tilde{J}})_{ab}
\psi_{\alpha b} ) . 
\label{sugraf}
\end{equation}
Up until now, our derivation is valid for an
arbitrary harmonic function $h$.  Choosing $h$ of Sec.~2.1
corresponding to the asymptotically locally $AdS_4$ background
geometry, we find that the action (\ref{sugraf}) is
identical to the matrix theory 
effective action $\Gamma^{(0)}
+ \Gamma^{(1)}_B + \Gamma^{(1)}_{\rm spin-orbit}$
from Eqs.~(\ref{taction}), (\ref{symeff}) and (\ref{symso}). 
  
\section{Discussions}

Our analysis in this paper suggests that the supersymmetry 
might be the key element for the agreement between the matrix 
theory and the supergravity.  With sixteen supercharges,
the $F^4$ term in the supersymmetric Yang-Mills theory
effective action is strongly constrained to be determined
up to an overall numerical factor, which can in turn
be uniquely fixed by the known perturbative 
analysis of, for example, Ref.~\cite{joe}.
On the supergravity side, the bosonic background geometries
are determined by the BPS equations.  Once this
background geometry is determined, the fermionic parts
of the effective action can also be determined by the
supersymmetry.  Therefore, considering our previous
work \cite{hks2} that showed the agreement of the 
bosonic effective
action between the two approaches, it is not surprising
to find a precise agreement for the spin-orbit coupling
terms. 

A pleasing feature of the effective action Eq.~(\ref{sugraf})
is that as soon as we choose the background geometry 
satisfying the BPS ansatz (thereby requiring $h$ be a harmonic
function), the classical fermionic action from supergravity 
immediately assumes the form of the fermionic terms
generated by the supersymmetric completion of the bosonic
four derivative $F^4$ terms of the SYM theory.  
Furthermore, for an arbitrary harmonic function
$h$, the quadratic (free field) classical action ${\cal L}^{(2)}$,
Eq.~(\ref{gamma2}), looks as if it is a theory on a flat 
background geometry (including fermion term).  This behavior
is consistent with the flatness of the (2+1)-dimensional
SYM theory moduli space.
A similar behavior, in the context of the Yang-Mills 
quantum mechanics with sixteen supercharges, was observed
for the quadratic supersymmetric Yang-Mills theory 
effective action \cite{sethi2}, where the non-renormalization
theorem for the terms was also proved.  

The precise agreement between the matrix theory side 
description and supergravity was verified for an
arbitrary value of the longitudinal eleventh
circle size and for all distance $r$, consistent with
the DLCQ procedure of Ref.~\cite{susskind}, which
was conjectured to be valid for the finite $N$ (for a 
fixed value of $p_- = N/R$, $N$ is proportional
to $R$).  The background 
metric that produces this agreement is that of the 
asymptotically locally $AdS_4$ metric.  
In the decompactification 
limit of the eleventh circle, this background geometry  
reduces to that of $AdS_4 \times S^7$.  In this case, the
harmonic function $h$ vanishes like $r^{-6}$
as one approaches the asymptotic infinity, unlike
the asymptotically flat geometries where $h$ goes to one.  
It is amusing to note that, therefore, the relationship
Eq.~(\ref{ftrans}) between $\theta$ and $\psi$ is the 
multiplication by an infinitely large scale factor.  
This transformation
is rather similar to the `removal of the pole contribution'
for the spinning fields in the treatment of the AdS/CFT
correspondence, which yields the holographic identification
of the bulk fields and the boundary fields {\em up to} 
conformal transformation \cite{witten}.    

There are several lines of generalizations to the analysis
presented in this paper.  One issue is the determination
of the static potential between two membranes.  Eight fermion
terms of the (2+1)-dimensional SYM theory was, as noted before,
already non-perturbatively obtained in Ref.~\cite{sethi}.
On the supergravity side, the full expansion up to all
fermion terms of the superfield in terms of the component
fields is available in Ref.~\cite{dpp2}, at least in
the $AdS_4 \times S^7$ background geometry.  It will be 
interesting to explicitly verify if the agreement between 
the strong coupling SYM theory and the membrane dynamics 
in $AdS_4$ supergravity holds for eight fermion terms
and to test if, of the possible $256 \times 256$
membrane-membrane polarization states, only 256 states
have the vanishing static potentials.  Secondly, since
we expect that the consideration of the spinless probe
in the presence of a spinning source will produce the
same answer to the one obtained here, due to the two-body
nature of the source-probe dynamics, it will be interesting
to do the explicit calculations of the bosonic probe
action in the presence of a non-trivial gravitino field.
In this case, as noted in \cite{vijay}, the non-vanishing
gravitino field induces rotations in the background
geometry.  This was in fact an approach taken by \cite{kraus2}
for the supergravity side analysis to determine the spin-orbit 
couplings for particle dynamics, which was in turn shown to
be identical to that of the supersymmetric
Yang-Mills quantum mechanics
two fermion terms.   

\begin{center}
{\bf Acknowledgements}
\end{center}

We would like to thank Sangmin Lee for useful discussions.

\end{document}